\documentclass[conference]{IEEEtran}\IEEEoverridecommandlockouts

\usepackage{cite}

\usepackage{float}
\usepackage{latexsym}
\usepackage{amssymb}
\usepackage{amsmath}
\usepackage{color}
\usepackage{subcaption}

\usepackage{graphicx}
\usepackage{epstopdf}
\usepackage{caption}
\usepackage{algorithmic}
\usepackage{algorithm}
\usepackage{array}
\usepackage{dsfont}
\usepackage[font={footnotesize}]{caption}

\begin{document}
\title{Trust-Aware Network Utility Optimization in Multihop Wireless Networks with Delay Constraints}
%\author{Evripidis~Paraskevas,~\IEEEmembership{Member,~IEEE, }
%Tao~Jiang,~\IEEEmembership{Member,~IEEE, }
%        Punyaslok~Purkayastha,~\IEEEmembership{Member,~IEEE, } 
%             and~John~S.~Baras,~\IEEEmembership{Life Fellow,~IEEE}}% <-this % stops a space
\author{\IEEEauthorblockN{Evripidis Paraskevas\IEEEauthorrefmark{1},
Tao Jiang\IEEEauthorrefmark{2} and
%Punyaslok~Purkayastha\IEEEauthorrefmark{3}  and
John S. Baras\IEEEauthorrefmark{1}}
\IEEEauthorblockA{\IEEEauthorrefmark{1}Department of Electrical \& Computer Eng. and Institute for Systems Research,\\
University of Maryland, College Park, MD, USA \\ \IEEEauthorrefmark{2}Samsung Research America-Dallas, Richardson, TX, USA %\\\IEEEauthorrefmark{3}Qualcomm Technologies Inc., San Diego, CA, USA 
\\ Email: \{evripar, baras\}@umd.edu, t.jiang@sta.samsung.com}}%, punyaslo@qti.qualcomm.com}}
%\thanks{Evripidis Paraskevas, Ta is with the Department
%of Electrical and Computer Engineering, Georgia Institute of Technology, Atlanta,
%\thanks{Manuscript received April 19, 2005; revised September 17, 2014.}}

\maketitle

\begin{abstract}
Many resource allocation problems can be formulated as a constrained maximization of a utility function. Network Utility Maximization (NUM) applies optimization techniques to achieve decomposition by duality or the primal-dual method. Several important problems, for example joint source rate control, routing, and scheduling design, can be optimized by using this framework. 
In this work, we introduce an important network security concept, ``trust'', into the NUM formulation and we integrate nodes' trust values in the optimization framework. These trust values are based on the interaction history between network entities and community based monitoring. Our objective is to avoid routing packets though paths with large percentage of malicious nodes. We also add end-to-end delay constraints for each of the traffic flows. The delay constraints are introduced to capture the quality of service (QoS) requirements imposed to each traffic flow.
\end{abstract}


% \fi
%\IEEEpeerreviewmaketitle

\section{Introduction\label{sec:intro}}
The problem of resource allocation in wireless networks has been a growing area of research over the past decade. Recent advances in the area of network utility maximization (NUM) driven cross-layer design\cite{Chiang07},\cite{ChiangINFOCOM06},\cite{StaiTON} have led to efforts on top-down development of next generation wireless network architectures. By linking decomposition of the NUM problem to different layers of the network stack, we are able to design protocols, based on the optimal NUM derived algorithms, which provide much better performance gain over the current network protocols.

% Optimization-based approaches have been extensively used over the past several years to study resource allocation problems in communication networks, for instance congestion control by Kelly\cite{Kelly98}, Network Utility Maximization (NUM)\cite{Chiang07}.
%There are new challenges to apply such approaches to wireless networks comparing to wireline networks. Wireless spectrum is scarce, thus it is important to use the wireless channel efficiently. 
Traditionally, network protocols are strictly layered. Source rate control, routing and scheduling (e.g. back-pressure scheduling\cite{Tassiulas92}) are implemented independently at different layers. In order to achieve high performance and efficient resource utilization, these protocols should be jointly designed while the layered structure is preserved. 
%Furthermore, in wireline networks, links are disjoint with fixed capacities. 
However, the nature of wireless multihop networks imposes new challenges to this cross-layer design, since the wireless channel is a shared medium where the transmissions of users interfere with each other. The channel capacity is ``elastic'' (time-varying) and the contention over such shared and limited network resources provides a fundamental constraint for resource allocation. All these challenges cause interdependencies across users and network layers. In spite of these difficulties, there have been significant developments in optimization-based approaches that result in loosely coupled cross-layer solutions\cite{Lin06}.

%For multi-hop wireless networks, it is critical to perform resource control in a distributed manner. Because of the interference among wireless channels, a scheduling policy has to resolve the contention between various attempting transmissions, which require global information. For instance, the back-pressure scheduler\cite{Tassiulas92} is a centralized algorithm. Centralized algorithms require frequent information exchange among nodes, which is computationally complex and not feasible for multi-hop wireless networks. In this work, we propose a distributed scheduling algorithm, which is sub-optimal and achieves fair channel access. 

In recent years, network security has become increasingly important in the context of wireless multihop networks. Different types of network attacks can be released and affect significantly their performance.  In our work, we consider that the adversary is capable of releasing some form of \textit{denial of service} (DoS) attack. Hence, without proper security consideration, the network operation is possible to be disrupted. To capture the notion of security, we use ``trust weights''\cite{Theodorakopoulos06} in the network utility optimization process. These weights indicate whether a network entity (node) is malicious or not, based on its interactions with the other network entities. Thus, by using them, we enhance the correct operation of the network and its resilience to attacks.
%¨C whether they are joint MAC-routing or joint physical-MAC-routing optimizations it does not matter, the idea applies to all these problems. 
The trust weights are developed by our network community based on monitoring and are disseminated via efficient methods so that they are timely available to all nodes that need them~\cite{Marti00}.  

End-to-end delay is a critical \textit{quality of service} (QoS) requirement for resource-constrained wireless networks. Network applications, served from different traffic flows in the wireless network, have different delay requirements. For example, video streaming applications are time-critical and have strict delay requirements. Hence, it is crucial to take into account these delay constraints, corresponding to different classes of traffic flows, to our trust-aware NUM problem.

%We plan to use both network coding and swarm intelligence based techniques to accomplish this efficient and timely dissemination of trust values (they are equivalent to reputations) based on recent work by Baras [***]. 

In this paper, we incorporate the notion of security into the NUM problem, by using the trust values of the network nodes. Users get higher utility, when they relay packets though trusted paths. Hence, our proposed \textit{trust-aware NUM} process ensures that untrusted paths (with malicious entities) will not receive high traffic rate. We also add end-to-end delay constraints in the NUM problem based on \cite{Delay-Constraints-NUM}. These delay constraints indicate the QoS requirements of the different traffic flows. The notion of \textit{link capacity margin} \cite{Delay-Constraints-NUM} is used to control the end-to-end delay. Finally, we propose a distributed cross-layer optimization algorithm for the trust-aware NUM problem with delay constraints. The distributed algorithm is based on the dual decomposition into source rate control, average end-to-end delay control and scheduling subproblems. 
%We consider a set of flows that share the resources of a fixed wireless network. Each flow is described by its resource-destination node pair, with no \textit{a priori} established routes. The effect of these trust weights on the resulting protocols is that in the scheduling problems (whether they are at the MAC or the routing protocol) the trustworthiness of the node will be automatically be considered and used. For example packets will not be routed as frequently to suspicious nodes. Or suspicious nodes will not be scheduled by the MAC protocol.

%We present decentralized algorithms which obtain a set of feasible solutions for the NUM problem, and compare the decentralized solutions with results of a centralized algorithm which is optimal and solves the NUM problem. Furthermore, we compare our results with schemes which do not take trust into consideration.

The rest of this paper is organized as follows. Section~\ref{sec:related} reviews the related work in the literature on network utility maximization (NUM) problem formulation and its security considerations. 
Section~\ref{sec:model} introduces the system model that we consider in this paper, including the network model, the adversary model, the trust values estimation, and the interference model. Section~\ref{sec:NUMFormulation} outlines the optimization constraints, which include link capacity, average end-to-end delay, and scheduling constraints, as well as the primal optimization problem. The dual function and its decomposition into different subproblems is studied in Section~\ref{sec:decomp}. Section~\ref{sec:distAlg} discusses the distributed algorithm for solving the network utility maximization (NUM) problem. The simulation results for our proposed trust-aware NUM problem with delay constraints are shown in Section~\ref{sec:eval}. Section~\ref{sec:conc} concludes this paper and discusses future work.

\section{Related Work\label{sec:related}} 

Network utility maximization (NUM) problems have been investigated widely during recent years. Most of works~\cite{Chiang07}, \cite{ChiangINFOCOM06}, \cite{StaiTON}, \cite{Lin06} focus on using NUM for cross-layer optimization. Chiang et. al~\cite{Chiang07} introduced a methodology for optimizing functional modules of the network, such as congestion control, routing and scheduling, through optimization decomposition. Chen et. al~\cite{ChiangINFOCOM06} proposed a subgradient algorithm for cross-layer optimization and its extension to time-varying channels and adaptive multi-rate devices. The proposed solution in most of these works depends on the decomposition of the dual function to different subproblems. Decomposition methods for NUM problem are proposed in \cite{Palomar06}.

Several works have introduced delay considerations for the traffic flows into the NUM problem formulation. Trichakis et. al~\cite{Trichakis08dynamicnetwork} proposed a dynamic NUM formulation with delivery contracts for the different traffic flows. Delivery contracts ensure that some quantity of a traffic flow will be delivered during a time interval. One other concept for delay, used for the NUM problem, is the \textit{link capacity margins}, which we use in our work. These margins were introduced in~\cite{Delay-Constraints-NUM} and~\cite{DBLP:journals/corr/HajiesmailiTK15} to control the average end-end delay. Link margins represent the estimated delay of the link, because higher link margin indicates lower link congestion and thus less delay.

As far as we are concerned, there are not a lot of works that relate security with the NUM problem~\cite{TaoISCCSP},~\cite{Jamming-Aware-NUM}. Tague et. al~\cite{Jamming-Aware-NUM} proposed a jamming-aware throughput maximization approach. The authors estimate the effect of jamming on packet delivery ratio. Then, they use these jamming estimates in the NUM problem to allocate data traffic appropriately in order to achieve throughput maximization. They adopt an objective function, based on portfolio selection theory to maximize throughput for the different source nodes.

To the best of our knowledge, our work is the first to study trust-aware network utility maximization problems. Trust values affect the outcome of the NUM process and make it resilient to malicious nodes' behavior.

\section{System Model\label{sec:model}}

\subsection{Network Model}
We consider a multihop wireless network that can be defined by a graph $G(\mathcal{N},\mathcal{L})$. The vertex set $\mathcal{N}$ represents the wireless network nodes. The edge set $\mathcal{L}$ represents the wireless links. An ordered pair of nodes $(i,j)$ belongs to the edge set $\mathcal{L}$ if and only if node $j$ can receive data packets directly from node $i$. For simplicity, we also use the symbol $\ell$ to denote a wireless link. We assume that all node-to-node communication is unicast, i.e. each packet transmitted by a node $i\in{\mathcal{N}}$ is intended for a unique $j\in{\mathcal{N}}$ with $(i,j)\in\mathcal{L}$. Each of the wireless links has a maximum capacity denoted by $c_{i,j}$. The interference constraints among transmission links will be described in a later subsection. 

\begin{figure}[!t]
\begin{center}
\includegraphics[width=0.35\textwidth]{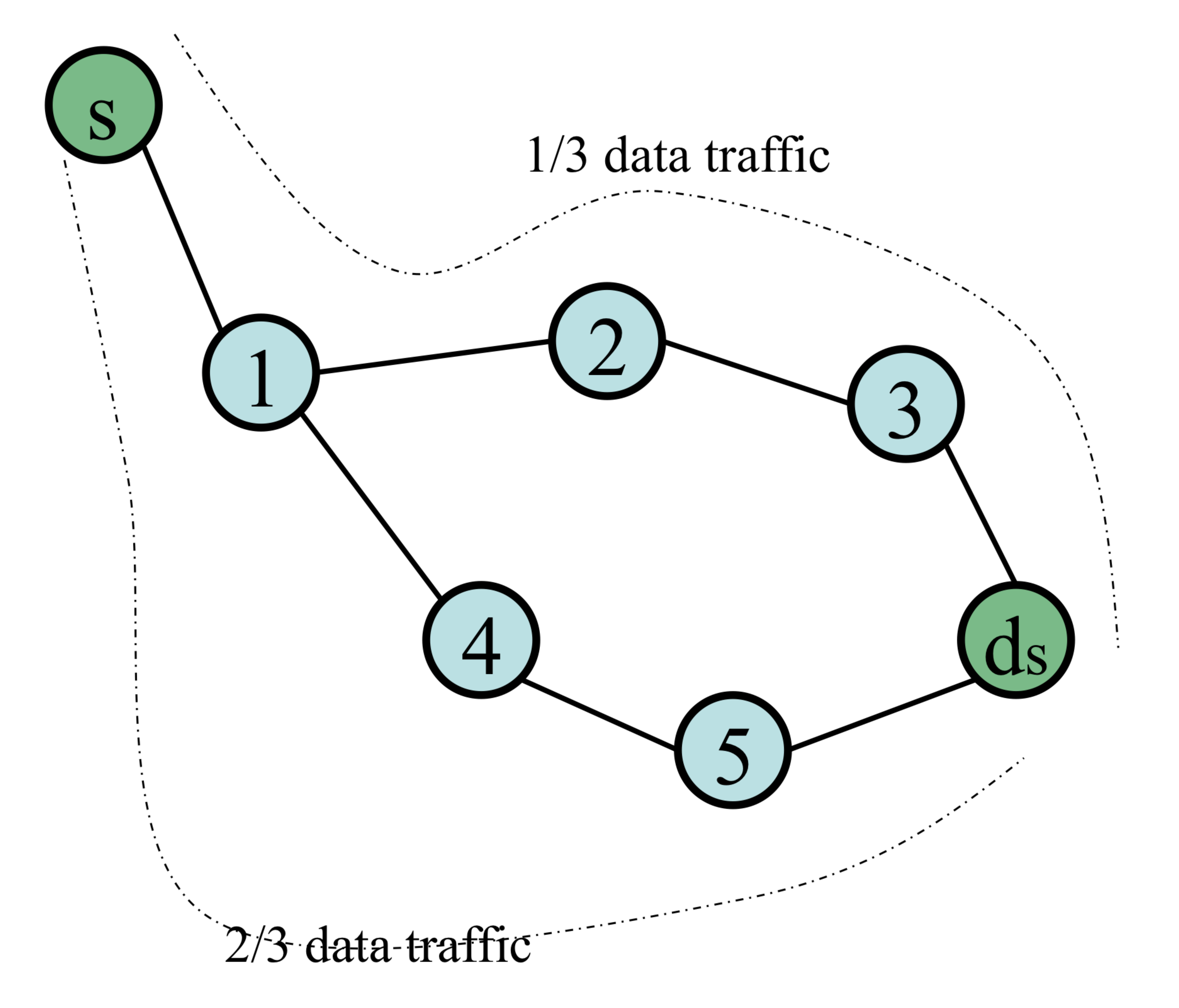}
\caption{Network with two alternative paths for traffic flow from $s$ to $d$.\label{fig:flow_trust}}
\end{center}
\end{figure}

There is a set $\mathcal{F}$ of network traffic flows that share the wireless network resources and each flow $f\in{\mathcal{F}}$ is associated with a source node $s$. Each source node $s$ in a subset $\mathcal{S}\subseteq{\mathcal{N}}$ generates data packets for a single destination node $d_{s} \in \mathcal{N}$. We assume that each source node $s$ constructs multiple routing paths with multiple hops to $d_{s}$ in order to distribute the traffic demand and satisfy the flow related QoS requirements. We denote as $\mathcal{P}_{s} = \{p_{s1},\ldots,p_{sP_{s}}\}$ the collection of the alternative paths $P_{s}$ that can be used to route packets from $s$ to $d_{s}$. Each path $p_{sk}\in{\mathcal{P}_{s}}$ is specified by a subset of wireless links and is assumed to be loop-free. 
An example of two different paths $p_{s1}$ and $p_{s2}$ from $s$ to $d_{s}$ is shown in Figure~\ref{fig:flow_trust}, where
\begin{align*}
	p_{s1} & = \{(s,1), (1,2), (2,3), (3,d_{s})\} \\
	p_{s2} &=  \{(s,1), (1,4), (4,5), (5,d_{s})\} 
\end{align*}
%\vspace{-5pt}
%\vspace{-10pt}
Let $\mathbf{x_{s}}$ denote the $P_{s}\times{1}$ traffic rate vector with which data packets are sent from $s$ to $d_{s}$ over multiple paths $p_{sk}\in{\mathcal{P}_{s}}$, and multiple hops. Each component of the vector $x_{sk}$ denotes the proportion of traffic rate allocated to the corresponding path $p_{sk}$, which routes data packets from source node $s$ to destination $d_{s}$. The total data rate of the source $s$ is given by the summation of $x_{sk}$ over $k=1,\ldots,{P_{s}}$.

We assume that the traffic rate vector $\mathbf{x_{s}}$ of each flow is constrained to a non-negative orthant. The traffic rate allocated to each traffic flow should also not exceed a maximum data rate $\mathcal{R}_{s}$. Therefore, each of the traffic rate vectors $\mathbf{x_{s}}$ should satisfy the following constraints
\begin{align}
	\mathbf{x_{s}} \geq \mathbf{0} & \hspace{10pt} ,\forall{s \in \mathcal{S}} \label{eq:Source-nonnegative}\\
	\mathbf{1}^T\mathbf{x_{s}} \leq{\mathcal{R}_{s}} & \hspace{10pt} ,\forall{s \in \mathcal{S}}
\end{align}
In Eq.~\eqref{eq:Source-nonnegative} each component of the vector is nonnegative.
These constraints define the convex set $\mathcal{X}_{s}$ of feasible traffic rate vectors $\mathbf{x_{s}}$ for source node $s$.

We denote by $\mathbf{R_{s}} = [(R_{s})_{k(i,j)}]_{(P_{s}\times|\mathcal{L}|)}$ the routing matrix that indicates the different paths from source node $s$ to destination $d_{s}$. Element $(R_{s})_{k(i,j)}$ of the routing matrix is defined as follows
\begin{equation}\label{eq:routing-matrix}
  (R_{s})_{k(i,j)}=\begin{cases}
    1, & \text{if $p_{sk}$ passes through link $(i,j)$}.\\
    0, & \text{otherwise}.
  \end{cases}
\end{equation}

\subsection{Security Considerations: Adversary Model and Trust}
\label{sec:SecurityModel}
In this paper, we study the network utility optimization problem with considerations of network security. All previous works on the network utility maximization (NUM) problems assume that nodes operate correctly. For example, intermediate nodes successfully forward all packets,  and they follow the routing and scheduling protocols. However, nodes do not always function correctly in reality. They may be compromised by attackers, their communication may be blocked or interfered by attackers, or they may just be misconfigured. Wireless networks are especially vulnerable to attacks because of the inherent properties of the shared wireless medium. Therefore, we believe it is crucial to take the security aspect into consideration in the NUM problems. We are going to define the adversary model, which describes the capabilities of an adversarial node, and the notion of trust, which indicates whether a node can be considered as trustworthy based on its observed behavior.

\par \textbf{\textit{Adversary Model:}} 
We assume that the adversarial node is not following the network protocol and attempts to disrupt communication by dropping or modifying data packets. In this work, we mainly consider that the adversary is capable of dropping data packets in a deterministic or probabilistic way. This type of attack leads to lack of \textbf{availability} of the network and constitutes a \textit{denial of service} (DoS) attack. The DoS attack affects significantly some QoS requirements, such as end-to-end delay and packet delivery ratio. Thus, in order to support \textit{time-critical} applications, the traffic allocation mechanisms should be resilient to these types of attacks. In general, the notion of trust can also address different types of attacks, such as the modification or fabrication of data messages. In this case, trust evaluation should incorporate authentication or inspection (filtering) mechanisms (e.g. Message Authentication Codes (MAC) or Deep Packet Inspection (DPI)~\cite{Sherry:2015:BDP:2785956.2787502}) at the receiver and intermediate nodes, in order to define the trustworthiness of a network node.

\par \textbf{\textit{Trust Estimates:}} 
The concept of security, which we adopt to distinguish misbehaving nodes in this work is \textbf{trust}.
Trust is a very critical concept not only in computer networks, but also in various other networks that involve intelligent decisions, such as social networks. All the connections and communications in these networks imply the existence of trust. Trust integrates with several components of network management, such as risk management, access control and authentication. 

Trust management is to collect, analyze and present trust related evidence and to make assessments and decisions regarding trust relationships between entities in a network~\cite{Blaze99}. The collection of trust evidence and the decision of trust are beyond the scope of this paper. We assume that there are mechanisms to efficiently distribute trust evidence, such that duplicates of evidence documents are  stored in places where they are most needed. Different approaches of trust evidence distribution could be found in \cite{Eschenauer02ontrust} and \cite{JiangGlobecom08} (use of \textit{network coding} to efficiently distribute trust credentials among network entities). Once the trust evidence is in hand, nodes could evaluate the trustworthiness of other nodes. For instance, in wireless environments, the monitoring mechanisms can help detect the behaviors of neighboring nodes and thus infer their trust values\cite{Marti00}. We define the trust estimated value (or trustworthiness) of node $i$ as $\nu_{i}$. 

There are various ways to represent trust values $\nu_{i}$ numerically. In different trust schemes, continuous or discrete numerical values are assigned to measure the level of trustworthiness for a network entity. For example, in~\cite{Maurer:1996:MPI:646646.699185} the entity's opinion about the trustworthiness of a digital certificate is defined as a continuous value in $[0,1]$. Using the same logic for our definition of trust, we denote that it takes a continuous numerical value in $[0,1]$.

We define an \textit{update period} of the trust estimates denoted by $T_{update}$. During the update period, represented by the time interval $[t-T_{update},t]$, the trust evaluation mechanism provides fresh estimates of the trust values for nodes $i \in \mathcal{S}$, based on the interaction between network entities. Each node evaluates trust estimates for its neighbor nodes and then the trust mechanism propagates the trust estimates throughout the network. Hence, at the time that we need to transmit data packets, we use the trust estimates derived at the latest update period.

In order to prevent significant variation in the trust estimate $\nu_{i}$ of node $i$ and to include memory of the trust evaluation, we suggest using an exponential
weighted moving average (EWMA)~\cite{doi:10.1080/00401706.2000.10485986} to update the trust estimate as a function of the previous estimate, as indicated in~\cite{Jamming-Aware-NUM}. Hence, the trust value of node $i$ at time $t$ is given by
\begin{equation}\label{eq:EWMA-trust}
	\nu_{i}(t) = (1-\alpha){\nu_{i}(t-T_{update})} + \alpha \nu_{i}^{new},
\end{equation}
where $\alpha \in [0,1]$ is a constant weight indicating the relative preference between updated and historic samples of trust values and $\nu_{i}^{new}$ is the fresh estimate of trust value for node $i$, given from the trust evaluation mechanism.

Given the trust values for the intermediate nodes across a path $p_{sk}$, the source node $s$ evaluates the updated \textit{aggregate trust value} for the path $p_{sk}\in \mathcal{P}_{s}$. The \textit{aggregate trust value} of the path $p_{sk}$ is denoted by $t_{sk}$ and can be expressed as the product of the corresponding trust values along the path as follows

\begin{equation}\label{eq:trust-path}
	t_{sk} = \prod_{j:(i,j)\in{p_{sk}}} \nu_{j}
\end{equation}
Source nodes evaluate the aggregate trust values for their alternative routing paths to destination $d_{s}$, in order to determine the optimal traffic allocation among the different paths. 

One additional parameter that we should consider in the data traffic allocation process is \textit{path reliability}~\cite{Rate-ReliabilityTradeoff}. In our work, path reliability is indicated by the corresponding aggregate trust value $t_{sk}$ over the path $p_{sk}$, which denotes the proportion of the allocated traffic flow that is actually received at destination node $d_{s}$. Hence, in order to maintain the reliability of the network the received traffic rate for each traffic flow should exceed a certain threshold. We denote this threshold for each source node $s$ as $\mathcal{R}^{thres}_{s}$, which is proportional to the maximum allowable rate $\mathcal{R}_{s}$. Thus, our allocated traffic rate for each source node should satisfy the reliability constraint expressed as
\begin{equation}
	\sum_{p_{sk} \in \mathcal{P}_{s}} t_{sk}x_{sk} \geq{\mathcal{R}^{thres}_{s}}, \hspace{10pt} \forall{s \in \mathcal{S}}
\end{equation}
The convex set $\mathcal{X}_{s}$ of feasible traffic rate vectors for source node $s$ should also satisfy the above reliability constraint.

\subsection{Interference model and capacity region}\label{sec:capacity-region}

In this subsection, we describe the interference model and the feasible capacity region. In order to model interference among wireless links of our original wireless multihop network, we use the concept of the \textit{conflict graph} introduced in~\cite{Jain:2003:IIM:938985.938993}. The conflict graph captures the contention relations among the links. Each vertex in the conflict graph indicates a wireless link and each edge indicates the interference between the two corresponding links.

We can detect all the independent sets of vertices in the conflict graph. We denote an independent set of links by $e$. The independent set $e$ can be represented as a $\mathcal{|L|}\times1$ capacity vector $r^{e}$. The element $r^{e}_{i,j}$ is expressed as
\begin{equation}\label{eq:independent-set}
  r^{e}_{i,j}=\begin{cases}
    c_{i,j}, & ,\text{if $(i,j)\in{e}$}.\\
    0 & ,\text{otherwise}.
  \end{cases}
\end{equation} 
The links that belong in an independent set do not interfere and are allowed to transmit simultaneously. The feasible capacity region $\Lambda$ \cite{ChiangINFOCOM06} is defined as the convex hull of these capacity vectors and is expressed as
\begin{equation}
\Lambda = \Big\{r: r = \sum_{e}{\beta_{e}r^{e}}, \beta_{e} \geq{0}, \sum_{e}{\beta_{e}} = 1\Big\}	
\end{equation}

Hence, the scheduling constraint indicates that the allocated capacity vector from the scheduling process, denoted by $\mathbf{\hat{c}}$, should satisfy $\mathbf{\hat{c}} \in \Lambda$.

\section{Network Utility Maximization Formulation}\label{sec:NUMFormulation}
In this section, we present the optimization framework for trust-aware network utility maximization (NUM), in the case that the network nodes have an updated estimate of trust values. We first develop a set of constraints imposed to our utility optimization problem. These constraints are related to capacity of wireless links, average end-to-end delay and scheduling. Then, we formulate the trust-aware utility optimization problem, which gives an optimal solution to the traffic flow allocation problem. 

\subsection{Optimization Constraints}

\par \textbf{\textit{Link Capacity constraint:}}
To define capacity constraints we first introduce the \textit{link capacity margin} optimization variables, which were initially introduced in~\cite{Delay-Constraints-NUM} and~\cite{DBLP:journals/corr/HajiesmailiTK15}, in order to capture the imposed delay constraints. We denote by $\sigma_{i,j}$ or simply $\sigma_{\ell}$, the link capacity margin of link $(i,j)\in\mathcal{L}$. Link capacity margin is defined as the difference between scheduled (allocated) capacity of a wireless link and the maximum allowable traffic flow passing though it. Link capacity margin is used to control link delay and therefore the average end-to-end delay.

We also need to take into account our trust estimates for the capacity constraints of each link $(i,j)\in{\mathcal{L}}$. Based on the capabilities of the malicious nodes, described in Section~\ref{sec:SecurityModel}, the initially allocated traffic rate $x_{sk}$ can be significantly reduced at malicious intermediate nodes because of dropping attacks. The decrease of the traffic rate is proportional to the \textit{aggregate trust value} of the selected path. To be more specific, the decrease of the rate observed at an intermediate node is proportional to the aggregate trust value up to this intermediate node. Let $p_{sk}^{(i,j)}$ denote the sub-path of $p_{sk}$ from source node $s$ to the intermediate node $j$ through link $(i,j) \in p_{sk}$. Then the traffic rate forwarded by intermediate node $j \in \mathcal{N}$ is computed by $t_{sk}^{(i,j)}x_{sk}$, where $t_{sk}^{(i,j)}$ is evaluated as the product of trust estimates over the sub-path $p_{sk}^{(i,j)}$, given by Eq.~\eqref{eq:trust-path}.

Hence, the capacity constraint associated with each wireless link $(i,j)\in{\mathcal{L}}$ is formulated as follows 
\begin{equation}\label{eq:CapacityConstraint}
\sum_{s\in\mathcal{S}}\sum_{k:(i,j)\in{p_{sk}}}{t_{sk}^{(i,j)}x_{sk}} \leq {\hat{c}}_{i,j} - \sigma_{i,j}, \hspace{8pt} \forall{(i,j)\in{\mathcal{L}}},
\end{equation}
where $\hat{c}_{i,j}$ is the capacity allocated to the wireless link $(i,j)\in{\mathcal{L}}$.

To define the different sub-paths' aggregate trust values, we denote by $\mathcal{T}_{s}$ the $P_{s}\times{|\mathcal{L}|}$ \textit{aggregate trust incidence matrix} for source $s$, with rows indexed by the alternative paths $p_{sk}$ and columns indexed by links $(i,j)$. If a link $(i,j)$ does not belong to any of the possible paths $p_{sk}$ for source $s$, then the corresponding entry of the incidence matrix is equal to $0$. The element $t(p_{sk}, (i,j))$ or otherwise $t_{sk}^{(i,j)}$ for row $p_{sk}$ and column $(i,j)$ of $\mathcal{T}_{s}$ denotes the aggregate trust value of a possible sub-path $p_{sk}^{(i,j)}$ of path $p_{sk}$ and is given by
 \begin{equation}\label{eq:aggtrust-matrix}
  t(p_{sk}, (i,j))=\begin{cases}
    \underset{j^{'}:(i^{'},j^{'})\in{p_{sk}^{(i,j)}}}\prod \nu_{j^{'}} & ,\text{if $(i,j)\in{p_{sk}}$}\\
    0 & ,\text{otherwise}
  \end{cases}
  \end{equation}

\par \textbf{\textit{Average end-to-end delay constraint:}}
By using the link margin variables $\sigma_{i,j}$, we define as $\phi(\sigma_{i,j})$ the delay of link $(i,j)\in{\mathcal{L}}$. The function $\phi(\cdot)$ is typically a strictly convex, nonnegative valued, function of $\sigma$. The packet arrival process model determines the way that $\phi(\cdot)$ depends on $\sigma_{i,j}$. As described in~\cite{Delay-Constraints-NUM} and \cite{DBLP:journals/corr/HajiesmailiTK15}, for Poisson process arrival, we have
\vspace{-5pt}
\begin{equation}\label{eq:delay-estimate}
	\phi(\sigma_{i,j}) =\phi_{i,j} = \frac{1}{\sigma_{i,j}}
\end{equation}
We define by $\phi(\sigma)$ the vector that has components the delay of all links of the network.
%We denote by $\mathbf{R_{s}} = [(R_{s})_{kl}]_{(P_{s}\times|\mathcal{L}|)}$ the routing matrix that indicates the different routing paths for source node $s$ to destination $d_{s}$. Element $(R_{s})_{kl}$ of the routing matrix is defined as follows:
%\begin{equation}
%  (R_{s})_{k\ell}=\begin{cases}
%    1, & \text{if $p_{sk}$ passes through link $\ell$}.\\
%    0, & \text{otherwise}.
%  \end{cases}
%\end{equation}

Delay constraints indicate the QoS requirements imposed to a specific traffic flow. Traffic flows that serve time-critical applications should have strict delay constraints. 
The end-to-end delay is expressed by adding the link delays for each of the links over path $p_{sk}$ of source node $s$. We denote the upper bound average delay constraint for each of the multiple paths of the source node $s$ as $\mathcal{D}_{s}>0$. Hence, we have that the average end-to-end delay constraint for every source node $s$ is given by (using the routing matrix expressed in Eq.~\eqref{eq:routing-matrix})
%\vspace{-1pt}
\begin{equation}\label{eq:avgDelay}
	\mathbf{R_{s}}{\bf{\phi(\sigma)}} \leq \mathbf{1}\mathcal{D}_{s}, \hspace{10pt} \forall{s\in{\mathcal{S}}}
\end{equation}

\par \textbf{\textit{Scheduling constraint:}}
The capacity $\hat{c}_{i,j}$ allocated to the wireless link $(i,j)$ should lie on the capacity region specified by $\Lambda$, which we describe in Sec.~\ref{sec:capacity-region}. Hence, our scheduling constraint is expressed as
\begin{equation}\label{eq:Scheduling}
	\mathbf{\hat{c}} \in \Lambda
\end{equation}

\subsection{Utility Optimization}
To determine the optimal traffic rate allocation to the different paths $\mathcal{P}_{s}$, each source $s$ chooses a utility function $\mathcal{U}_{s}(\cdot)$ that evaluates the total data rate delivered to the destination $d_{s}$. Utility functions $\mathcal{U}_{s}(\cdot)$ are chosen to be strictly concave, continuous, monotonically increasing and twice differentiable. 

Trust estimates for the different paths $p_{sk}$ of a source node (defined in Sec.~\ref{sec:SecurityModel}) should be incorporated to the selected utility function $\mathcal{U}_{s}(\cdot)$. Source nodes should obtain greater utility when they decide to allocate higher traffic rate through routing paths with higher aggregate trust value $t_{sk}$. Hence, the utility function for each source node $s\in \mathcal{S}$ can be selected as
\begin{equation}
 \mathcal{U}_{s}(\mathbf{x}_{s}) = \sum_{p_{sk} \in \mathcal{P}_{s}}\Big(t_{sk}\log{(x_{sk})}\Big)	
\end{equation}

The \textit{primal utility optimization problem} formulation, based on the capacity, average end-to-end delay and scheduling constraints described in Eq.~\eqref{eq:CapacityConstraint}, \eqref{eq:avgDelay} and~\eqref{eq:Scheduling}, is given by
%\begin{equation}\label{eq:primal}
%\begin{aligned}
%& \underset{\mathbf{x},\mathbf{\sigma},\mathbf{\hat{c}}}{\text{maximize}}
%& & \sum_{s\in\mathcal{S}}\mathcal{U}_{s}(\mathbf{x}_{s}) \\ 
%& \text{subject to}
%& &  \sum_{s\in\mathcal{S}}\sum_{k:(i,j)\in{p_{sk}}}{t_{sk}^{(i,j)}x_{sk}} \leq {\hat{c}}_{i,j} - \sigma_{i,j}, \ \forall{(i,j)\in{\mathcal{L}}}\\
%& & & \mathbf{R_{s}}{\bf{\phi(\sigma)}} \leq \mathbf{1}\mathcal{D}_{s}, \hspace{43pt} \forall{s\in{\mathcal{S}}} \\
%%& & \mathbf{x_{s}} \geq \mathbf{0} & \hspace{10pt} ,\forall{s \in \mathcal{S}}\\
%& & & 0\leq \mathbf{1}^{T}\mathbf{x_{s}} \leq{\mathcal{R}_{s}}, \hspace{41pt} \forall{s \in \mathcal{S}}\\
%%& & &0\leq \mathbf{1}^{T}\mathbf{x_{s}} \mathbf{t_{s}}\leq{R_{s}}, \hspace{12pt} \forall{s \in \mathcal{S}}\
%& & &\sum_{p_{sk} \in \mathcal{P}_{s}} t_{sk}x_{sk} \geq{\mathcal{R}^{thres}_{s}}, \hspace{10pt} \forall{s \in \mathcal{S}}\\
%& & & \ \mathbf{\hat{c}} \in \Lambda
%\end{aligned}
%\end{equation}
%\vspace{-5pt}
\begin{small}
 \begin{subequations}
\begin{eqnarray}
\underset{\mathbf{x},\mathbf{\sigma},\mathbf{\hat{c}}}{\text{max}} &&
 \sum_{s\in\mathcal{S}}\mathcal{U}_{s}(\mathbf{x}_{s}) \label{eq:primal}  \\ 
 %\begin{subequations}
 \text{s. t.} && 
  \sum_{s\in\mathcal{S}}\sum_{k:(i,j)\in{p_{sk}}}{t_{sk}^{(i,j)}x_{sk}} \leq {\hat{c}}_{i,j} - \sigma_{i,j}, \forall{(i,j)} \label{eq:capconstraint}\\ % \in{\mathcal{L}}}\\
 && \mathbf{R_{s}}{\bf{\phi(\sigma)}} \leq \mathbf{1}\mathcal{D}_{s}, \hspace{43pt} \forall{s\in{\mathcal{S}}}\label{eq:delayconstraint} \\
%& & \mathbf{x_{s}} \geq \mathbf{0} & \hspace{10pt} ,\forall{s \in \mathcal{S}}\\
 && 0\leq \mathbf{1}^{T}\mathbf{x_{s}} \leq{\mathcal{R}_{s}}, \hspace{41pt} \forall{s \in \mathcal{S}} \label{eq:rateconstraint}\\
%& & &0\leq \mathbf{1}^{T}\mathbf{x_{s}} \mathbf{t_{s}}\leq{R_{s}}, \hspace{12pt} \forall{s \in \mathcal{S}}\
&& \sum_{p_{sk} \in \mathcal{P}_{s}} t_{sk}x_{sk} \geq{\mathcal{R}^{thres}_{s}}, \hspace{10pt} \forall{s \in \mathcal{S}}\label{eq:reliabilityconstraint}\\
 && \mathbf{\hat{c}} \in \Lambda
\end{eqnarray}
 \end{subequations}
\end{small}
%\vspace{-10pt}
%\vspace{-10pt}
%\begin{eqnarray}
%\mbox{maximize}_{\mathbf{\hat{x}}} && \sum_{f}U_f(\hat{x}_f)  \\
% \textrm{subject to} & & \mathbf{x} \in \Lambda \label{eqn:const1}\\
% && \hat{x}_f = g_f\cdot x_f \textrm{ for all } f \in \mathcal{F} \label{eqn:const2}
%\end{eqnarray}

The trust-aware utility optimization problem is a strongly convex optimization problem, due to the strict concavity assumption of $U_{s}(\cdot)$ and the convexity of the capacity region. Therefore, there exists a unique optimal solution for the above primal problem, which we refer to as $(\mathbf{x^*},\mathbf{\sigma^*},\mathbf{\hat{c}^*})$. 
\section{Dual Decomposition Algorithm}\label{sec:decomp}
In this section, we solve the utility optimization problem described in Eq.~\eqref{eq:primal} by applying dual decomposition~\cite{Palomar06},~\cite{Boyd04}. The decomposition of the optimization problem provides distributed algorithms, which solve the underlying optimization problem. We note that strong duality holds for our optimization problem (duality gap is zero) and thus we can solve it through its dual function. 

%As we will see some of the optimal algorithms require centralized information and thus are not feasible to implement in a distributed way. We will propose random access schemes to fit into the decomposed sub-optimal problems in Sec.~\ref{sec:distAlg} and evaluate the effect of these approximations having on optimality in Sec.~\ref{sec:eval}.

%Notice that the variables $g_f$ and $\hat{x}_f$ are coupled by the second constraint Eq.~(\ref{eqn:const2}). In this work, we take the log of variables to decouple $g_f$ and $\hat{x}_f$ and log change of variables and constants: $\hat{x}'_f=\log \hat{x}_f$, $g'_f = \log g_f$, $x'_f = \log x_f$, and $U'_f(\hat{x}'_f)=U_f(e^{\hat{x}'_f})$. Now the primal problem is separable. Notice that by taking the log change, the problem may not be a convex optimization problem, since the objective $U'_f(\cdot)$ may not be a strictly concave function, even though $U_f(\cdot)$ is a strictly concave. However, the following simple sufficient condition guarantees its concavity:
%\begin{equation}
%\frac{\partial^2U_f(x)}{\partial x^2} < -\frac{\partial U_f(x)}{x\partial x} \label{eqn:curvature}
%\end{equation} 
%which states that the curvature (degree of concavity) of the utility function needs to be not just nonpositive but bounded away from zero by as much as $-(\partial U_f(x)/x\partial x)$, i.e., the application represented by this utility function must be elastic enough~\cite{Chiang07}. %(TEST THE UTILITY FUNCTION)

We define the Lagrange multipliers (dual variables) associated with the capacity and average end-to-end delay constraints. Let $\mathbf{\lambda}$ denote the $|\mathcal{L}|\times{1}$ vector of \textit{link prices} (dual variables) $\lambda_{i,j}$ (otherwise denoted by $\lambda_{\ell}$) associated with the capacity constraints for each wireless link. Also, let $\mathbf{\mu}_{s}$ denote the $P_{s}\times{1}$ vector of dual variables $\mu_{sk}$ associated with the average end-to-end delay constraints imposed to every traffic flow $s\in\mathcal{S}$.

In order to introduce the dual problem, we define the partial Lagrangian $L(\mathbf{x},\sigma,\hat{\textbf{c}},\lambda,\mu)$ of the optimization problem by using the inequality constraints given from Eq.~\eqref{eq:capconstraint} and~\eqref{eq:delayconstraint}
\vspace{-10pt}
%\begin{small}
\begin{multline}\label{eq:Lagrangian}
L(\mathbf{x},\sigma,\hat{\textbf{c}},\lambda,\mu)  = \\ \sum_{s\in\mathcal{S}}\mathcal{U}_{s}(\mathbf{x}_{s}) - \sum_{(i,j)\in\mathcal{L}}{\lambda_{i,j}}\Big(\sum_{s\in\mathcal{S}}\sum_{k:(i,j)\in{p_{sk}}}{\Big(t_{sk}^{(i,j)}x_{sk}\Big)} - {\hat{c}}_{i,j} +\sigma_{i,j}\Big) \\ - \sum_{s \in \mathcal{S}}{\mu_{s}^{T}\Big(\mathbf{R_{s}}{\bf{\phi(\sigma)}} -\mathbf{1}\mathcal{D}_{s}\Big)}  \\ 
=\sum_{s\in{\mathcal{S}}}{\Big(\mathcal{U}_{s}(\mathbf{x}_{s}) -\sum_{(i,j)\in\mathcal{L}}\lambda_{i,j}\sum_{k:(i,j)\in{p_{sk}}}{\Big(t_{sk}^{(i,j)}x_{sk}\Big)}}\Big) \\ -
\sum_{(i,j)\in\mathcal{L}}\Big(\phi_{i,j}{\Big(\sum_{s\in\mathcal{S}}\sum_{k:(i,j)\in{p_{sk}}}\mu_{sk}[(R_{s})_{k(i,j)}] \Big) + \lambda_{i,j}\sigma_{i,j}\Big)} \\ 
+\lambda^{T}\hat{\mathbf{c}} + \sum_{s\in{\mathcal{S}}}{\mu_{s}^{T}\mathbf{1}\mathcal{D}_{s}} = \\
 \sum_{s\in{\mathcal{S}}}{\Big(\mathcal{U}_{s}(\mathbf{x}_{s}) - (\lambda^{s})^{T}\mathcal{T}_{s}^{T}\mathbf{x}_s\Big)} - \sum_{(i,j)\in\mathcal{L}}\Big({\phi_{i,j}\mu^{(i,j)} + \lambda_{i,j}\sigma_{i,j}\Big)} \\
+\lambda^{T}\hat{\mathbf{c}} + \sum_{s\in{\mathcal{S}}}{\mu_{s}^{T}\mathbf{1}\mathcal{D}_{s}},
%+\max_{\mu \in \hat{\Gamma}}\sum_{(i,j)\in \mathcal{L}}\sum_{f\in \mathcal{F}}\mu^f_{ij}(\lambda_i^f-\lambda_j^fv_{j}) \\
%&& + \sum_f\max_{\mathbf{\hat{x}}' }\left\{U'_f(\hat{x}'_f)-\nu_f\hat{x}'_f\right\}. \label{eqn:dual}
\end{multline}
%\end{small}
where $\lambda^{s}$ is a sub-vector of the $\lambda$ dual variable and is associated with the constraint in Eq.~\eqref{eq:capconstraint}. It defines the $|\mathcal{L}|\times{1}$ column link price vector related to the links that belong to any of the paths $p_{sk}\in{\mathcal{P}_{s}}$ of a particular source node $s$ and is given by
 \begin{equation}\label{eq:lambda-source}
\lambda^{s}_{i,j}=\begin{cases}
    \lambda_{i,j} &, \text{if $(i,j)\in\underset{{p_{sk} \in \mathcal{P}_{s}}}\cup {p_{sk}}$}\\
    0 &, \text{otherwise}
  \end{cases}
  \end{equation}
and $\mu^{(i,j)} = \sum_{s\in\mathcal{S}}\sum_{k:(i,j)\in{p_{sk}}}\mu_{sk}[(R_{s})_{k(i,j)}]$  denotes the combination of dual variables $\mu$, which are related to a specific link $(i,j)$ and is associated with the constraint~\eqref{eq:delayconstraint}.

The dual objective function $h(\cdot)$ is then expressed as
%\vspace{-2pt}
%\begin{multline}\label{eq:dual}
%h(\mathbf{\lambda},\mathbf{\mu}) = \sup_{\mathbf{x} \in \mathcal{X}, \sigma \geq \mathbf{0}, \hat{\mathbf{c}} \in \Lambda}L(\mathbf{x},\sigma,\hat{\textbf{c}},\lambda,\mu) \\
%= \sup_{\mathbf{x} \in \mathcal{X}}\Big\{\sum_{s\in{\mathcal{S}}}{\Big(\mathcal{U}_{s}(\mathbf{x}_{s}) -  (\lambda^{s})^{T}\mathcal{T}_{s}^{T}\mathbf{x}_s\Big)}\Big\} \\ +\sup_{\sigma \geq \mathbf{0}}\Big\{-\sum_{(i,j)\in\mathcal{L}}\Big({\phi_{i,j}\mu^{(i,j)} + \lambda_{i,j}\sigma_{i,j}\Big)}\Big\} \\
%+\sup_{\hat{c} \in \Lambda} \Big\{\lambda^{T}\hat{\mathbf{c}}\Big\} + \sum_{s\in{\mathcal{S}}}{\mu_{s}^{T}\mathbf{1}\mathcal{D}_{s}}
%\end{multline}
\begin{small}
\begin{subequations}
\begin{eqnarray}
h(\mathbf{\lambda},\mathbf{\mu}) && =   %\sup_{\mathbf{x} \in \mathcal{X}, \sigma \geq \mathbf{0}, \hat{\mathbf{c}} \in \Lambda}L(\mathbf{x},\sigma,\hat{\textbf{c}},\lambda,\mu) \\
 \sup_{\mathbf{x} \in \mathcal{X}}\Big\{\sum_{s\in{\mathcal{S}}}{\Big(\mathcal{U}_{s}(\mathbf{x}_{s}) - (\lambda^{s})^{T}\mathcal{T}_{s}^{T}\mathbf{x}_s\Big)}\Big\}\label{eq:sourceratecontrol} \\ && +\sup_{\sigma \geq \mathbf{0}}\Big\{-\sum_{(i,j)\in\mathcal{L}}\Big({\phi_{i,j}\mu^{(i,j)} + \lambda_{i,j}\sigma_{i,j}\Big)}\Big\}\label{eq:end-to-enddelay} \\
&& +\sup_{\hat{c} \in \Lambda} \Big\{\lambda^{T}\hat{\mathbf{c}}\Big\}\label{eq:scheduling} \\
&& +\sum_{s\in{\mathcal{S}}}{\mu_{s}^{T}\mathbf{1}\mathcal{D}_{s}}
\end{eqnarray}
 \end{subequations}
 \end{small}

The dual optimization problem is defined by minimizing the dual objective function~\cite{Ber99} over the dual vector variables $\lambda$ and $\mu$ as follows
%\vspace{-1pt}
\begin{equation}\label{eq:dualopt}
\begin{aligned}
	& \underset{\mathbf{\lambda}\geq\mathbf{0},\mathbf{\mu}\geq\mathbf{0}}{\text{min}}
& &  h(\mathbf{\lambda},\mathbf{\mu})
\end{aligned}
\end{equation}

For given dual variables $\mathbf{\lambda}$ and $\mathbf{\mu}$, we can identify in the above equation of $h(\mathbf{\lambda},\mathbf{\mu})$ three decoupled maximization problems which we can solve separately. These three problems correspond to source rate control in Eq.~\eqref{eq:sourceratecontrol}, average end-to-end delay control in Eq.~\eqref{eq:end-to-enddelay}, and scheduling in Eq.~\eqref{eq:scheduling} respectively.

By solving these three independent optimization problems we can derive the optimal values for the primal optimization problem $\mathbf{x^*}(\mathbf{\lambda},\mathbf{\nu})$, $\mathbf{\sigma^*}(\mathbf{\lambda},\mathbf{\mu})$ and $\mathbf{\hat{c}^*}(\mathbf{\lambda},\mathbf{\mu})$  (described in Eq.~\eqref{eq:primal}). Given these values, we can then solve the dual problem by minimizing $h(\mathbf{\lambda},\mathbf{\mu})$ over $\mathbf{\lambda},\mathbf{\mu}\geq \mathbf{0}$.  There is no duality gap between the primal and the dual, because the capacity region $\Lambda$~\cite{ChiangINFOCOM06},~\cite{Jain:2003:IIM:938985.938993} is a convex set.

In the following subsections, we describe the decomposition of the dual objective function that leads to the cross-layer optimization problem and we specify the optimal solutions by solving these independent subproblems. 
%In the following subsections, we describe the decomposition of the dual objective function expressed in Eq.~\eqref{eq:dual} that leads to the cross-layer optimization problem and we specify the optimal solutions by solving the independent maximization subproblems. 
%In previous section, we derived the Lagrangian of the optimization problem as present in Eq.~(\ref{eqn:dual}). This section we decompose the problem accordingly and solve the independent maximization problems for data rate control, routing and scheduling.
\subsection{Source rate control}
Based on the dual decomposition the traffic rate vector of source node $s$ is determined by the first maximization subproblem in Eq.~\eqref{eq:sourceratecontrol}.
%\begin{equation}
%\hat{x}'_f=\mbox{argmax}_{x'^{\min}_f \leq x'_f\leq x'^{\max}_f}\sum_f \nu_fx'_f-\lambda_{s_f}^fe^{x'_f}. \label{eqn:rate_control}
%\end{equation}
%\begin{equation}
%\label{eq:rate_control}
%\mathbf{x_{s}^*} = \underset{\mathbf{x_{s}} \in \mathcal{X}_{s}}{\operatorname{argmax}}\Big\{\sum_{s\in{\mathcal{S}}}{\Big(\mathcal{U}_{s}(\mathbf{x}_{s}) - (\lambda^{s})^{T}\mathcal{T}_{s}^{T}\mathbf{x}_s\Big)}\Big\}
%\end{equation}
%\begin{equation*} 
%max_{x} \sum_{f} U_{f}({\color{red}{\nu_f}}x_{f})
%\end{equation*}
%It is important to note that each source node can adjust its data rate using 
%If the utility function $\mathcal{U}_{s}(\cdot)$ is invertible and its derivative is denoted by $\mathcal{U}_{s}^{'}(\cdot)$, then 
$U_{s}(\cdot)$ is a strictly concave scalar function of the rate vector variable $x_{s}$.
The maximization problem in~\eqref{eq:sourceratecontrol} is maximization of a concave function subject to the convex constraints~\eqref{eq:rateconstraint} and~\eqref{eq:reliabilityconstraint}. Thus, it has a unique solution.
$U_{s}(\cdot)$ is continuously differentiable.
Hence, the maximum will be given by the numerical solution of the equation  
%\begin{equation}
%\label{eq:rate_control_optimal}
%\mathbf{x_{s}^*} = \mathcal{U}_{s}^{'-1}\Big((\lambda^{s})^{T}\mathcal{T}_{s}^{T}\Big), \hspace{10pt} \forall {s\in\mathcal{S}}
%\end{equation}
%
%dUsubx/dx (the column vector representing the gradient of Usubs = scrptTsubs times lamda sups

%Each source node obtains an optimal value for its source rate expressed as
%\vspace{-5pt}
\begin{equation}
\label{eq:rate_control_optimal}
\nabla{\mathcal{U}_{s}}(x_{s}^*) = \mathcal{T}_{s}\lambda^{s}, \hspace{10pt} \forall {s\in\mathcal{S}}
\end{equation}
%\vspace{-5pt}
as long as the resulting solution for $x_{s}^*$ is in the interior of the constraint set defined by~\eqref{eq:rateconstraint} and~\eqref{eq:reliabilityconstraint}. Otherwise the solution will lie at the corners of the constrained set defined by \eqref{eq:rateconstraint} and~\eqref{eq:reliabilityconstraint}.
%The first subproblem is related to the source rate control for each source node of the wireless multi-hop network. Each source node attempts to adjust its traffic rate vector, in order to achieve higher utility. The configuration of the source rate control also depends on the estimated trust values of the different sub-paths of a specific source node. 
%
%Based on the dual decomposition the traffic rate vector of source node $s$ is determined by
%%\begin{equation}
%%\hat{x}'_f=\mbox{argmax}_{x'^{\min}_f \leq x'_f\leq x'^{\max}_f}\sum_f \nu_fx'_f-\lambda_{s_f}^fe^{x'_f}. \label{eqn:rate_control}
%%\end{equation}
%\begin{equation}
%\label{eq:rate_control}
%\mathbf{x_{s}^*} = \underset{\mathbf{x_{s}} \in \mathcal{X}_{s}}{\operatorname{argmax}}\Big\{\sum_{s\in{\mathcal{S}}}{\Big(\mathcal{U}_{s}(\mathbf{x}_{s}) - (\lambda^{s})^{T}\mathcal{T}_{s}^{T}\mathbf{x}_s\Big)}\Big\}
%\end{equation}
%
%%\begin{equation*} 
%%max_{x} \sum_{f} U_{f}({\color{red}{\nu_f}}x_{f})
%%\end{equation*}
%
%%It is important to note that each source node can adjust its data rate using 
%Therefore, if the utility function $\mathcal{U}_{s}(\cdot)$ is invertible and its derivative is denoted by $\mathcal{U}_{s}^{'}(\cdot)$, then each source node obtains an optimal value for its source rate
%
%\begin{equation}
%\label{eq:rate_control_optimal}
%\mathbf{x_{s}^*} = \mathcal{U}_{s}^{'-1}\Big((\lambda^{s})^{T}\mathcal{T}_{s}^{T}\Big), \hspace{10pt} \forall {s\in\mathcal{S}}
%\end{equation}

It is important to note that each source node $s$ is able to adjust its data rate vector using its local observations on the link prices $\lambda_{i,j}$ across the links of its multiple routing paths and the aggregate trust values of the different paths and their respective sub-paths.

\subsection{Average End-to-End Delay Control}
%\subsection{Average End-to-End Delay Control}
The second subproblem of the dual decomposition described in Eq.~(\ref{eq:end-to-enddelay}) is related to average end-to-end delay control based on the optimal values for the link capacity margin $\sigma_{i,j}$. 
%The optimal value for the capacity margin $\sigma_{i,j}$ of link $(i,j)$ is given by
%\begin{equation}
%\label{eq:delay_control}
%\mathbf{\sigma_{i,j}^*} = \underset{\sigma \geq \mathbf{0}}{\operatorname{argmax}}\Big\{-\sum_{(i,j)\in\mathcal{L}}\Big({\phi_{i,j}\mu^{(i,j)} + \lambda_{i,j}\sigma_{i,j}\Big)}\Big\}
%\end{equation}
%If the delay function $\phi(\cdot)$ is invertible and its derivative is denoted by $\mathcal{\phi}^{'}(\cdot)$, then there is 
%The optimal value for the link capacity margin of %each wireless link is obtained from the equation
%\vspace{-5pt}
Eq.~\eqref{eq:end-to-enddelay} is a strictly convex, minimization problem, subject to the constraint that all sigma are nonnegative. Thus, it has a unique solution. Function $\phi(\cdot)$ is a continuously differentiable function. Hence, the optimal values of $\sigma_{i,j}^*$  are obtained by solving the equations numerically

%\vspace{-3pt}
\begin{equation}
\label{eq:delay_control_opt}
\frac{d\mathcal{\phi}}{d\mathbf{\sigma}}(\sigma_{i,j}^*) \mu^{(i,j)} = - \lambda_{i,j}, \hspace{10pt} \forall {(i,j) \in\mathcal{L}}
\end{equation}

%The second subproblem of the dual decomposition is related to average end-to-end delay control based on the optimal values for the link capacity margin $\sigma_{i,j}$. This constraint is designed in order to satisfy the end-to-end delay requirements (QoS requirements) imposed for each traffic flow associated with a source node $s$.  
%
%Based on the dual decomposition, the optimal value for the capacity margin $\sigma_{i,j}$ of link $(i,j)$ is given based on Eq.~\eqref{eq:dualopt} by
%\begin{equation}
%\label{eq:delay_control}
%\mathbf{\sigma_{i,j}^*} = \underset{\sigma \geq \mathbf{0}}{\operatorname{argmax}}\Big\{-\sum_{(i,j)\in\mathcal{L}}\Big({\phi_{i,j}\mu^{(i,j)} + \lambda_{i,j}\sigma_{i,j}\Big)}\Big\}
%\end{equation}
%
%Hence, if the delay function $\phi(\cdot)$ is invertible and its derivative is denoted by $\mathcal{\phi}^{'}(\cdot)$, then there is an optimal value for the link capacity margin of each wireless link expressed as
%
%\begin{equation}
%\label{eq:delay_control_opt}
%\mathbf{\sigma_{i,j}^*} = \Big[\mathcal{\phi}^{'-1}\Big(-\frac{\lambda_{i,j}}{\mu^{(i,j)}}\Big)\Big]^{+}, \hspace{10pt} \forall {(i,j) \in\mathcal{L}}
%\end{equation}

By Eq.~\eqref{eq:delay_control_opt}, we observe that the updated dual variable related to the corresponding wireless link is needed. Thus, explicit/implicit exchange of dual variables between the different data sources is enabled.

%\subsection{Route selection}
%The second term of the dual function Eq.~(\ref{eqn:dual}) determines the route. 
%\begin{equation}
%\mathbf{g'}=\mbox{argmax}_{\mathbf{g'}}\sum_f \nu_fg'_f. \label{eqn:argmax_g}
%\end{equation}
%For the set of routes $R_f$, we define 
%\begin{equation}
%v_{\max}^f = \max_{r \in R_f} \prod_{(i,j) \in r}v_{j}
%\end{equation}
%\begin{equation}
%r_{\max}^f = \mbox{argmax}_{r \in R_f} \prod_{(i,j) \in r}v_{j}
%\end{equation}
%We have that 
%\begin{equation}
%g^*_f=v_{\max}^f. \label{eqn:argmax_g_s}
%\end{equation}
%Therefore, the optimal route for flow $f$ is the route with the highest trust value product. The modified Dijkstra's algorithm can be used to calculate the optimal route. %(GIVE THE DETAIL)

\subsection{Scheduling policy}
The third problem of the dual decomposition determines the scheduling policy. The optimal value for the allocated link capacity $\hat{c}_{i,j}^{*}$ is given by Eq.~\eqref{eq:scheduling}

%\vspace{-3pt}
\begin{equation}\label{eq:scheduling_opt}
	\hat{c}_{i,j}^{*} = \underset{\hat{c}_{i,j}\in{\Lambda}}{\operatorname{argmax}}{\sum_{(i,j)\in\mathcal{L}} {\lambda_{i,j}\hat{c}_{i,j}}}
\end{equation}

We need to find a scheduling policy so that the aggregate link weight $\sum_{(i,j)\in\mathcal{L}} {\lambda_{i,j}\hat{c}_{i,j}}$ could be maximized. The solution to this scheduling subproblem is based on the \textit{maximum weight} scheduling policy introduced in~\cite{Delay-Constraints-NUM}. This policy is described in Alg.~\ref{alg:scheduling}.

\begin{algorithm}
\caption{Maximum-Weight Scheduling Policy~\cite{Delay-Constraints-NUM}
}\label{alg:scheduling}
\begin{algorithmic}[1]
%\STATE Node $i$ performs scheduling.
\STATE Start scheduling timer
\WHILE{Timer is ON}
\IF {$i$ has been scheduled by any of its neighboring nodes}
\STATE Send messages to notify all nodes in its interference set and stop the process.
\ELSE 
\STATE Each incident link $(i,j)$ is assigned a weight $\lambda_{i,j}\hat{c}_{i,j}$. 
\STATE Find a link with the maximum weight among all incident links.
\IF {link $(i,j)$ is found}
\STATE Schedule the link with the effective allocated capacity $\hat{c}_{i,j}$. 
\STATE Notify all neighbors in the interference set.
\ELSE 
\STATE Stop the process
\ENDIF
\ENDIF
\ENDWHILE
\end{algorithmic}
\end{algorithm}

\subsection{Distributed Algorithm\label{sec:distAlg}}
In this section, we describe the distributed algorithm that solves the network utility optimization problem. Our solution is based on subgradient descent iterative methods for the update of the dual variables. We first compute the subgradient of the dual function with respect to each of the dual variables and then we propose our distributed algorithm.

%Scheduling sub-problem discussed in the last section requires global knowledge on the rate vector, which becomes the bottleneck for solutions in mobile ad hoc networks. In the this section, we study the distributed implementation of the maximization problem. We first start with a centralized algorithm which converges to the global optimum with time.

%We assume that time is divided into slots. At each time slot, source nodes choose the traffic rate vector and the scheduling policy will select data to be forwarded on each link. 

%The source node of each flow uses its local multiplier and the utility function associated with that flow to update the flow rate in an iterative manner. One example of the rate controller is directly derived from Eq.~(\ref{eqn:rate_control}):
%\begin{equation}
%\hat{x}'_f[t+1]=\mbox{argmax}_{x'^{\min}_f \leq x'_f\leq x'^{\max}_f}\sum_f \nu_f[t]x'_f-\lambda_{s_f}^f[t]e^{x'_f}. \label{eqn:rate_control_iterate}
%\end{equation}
%Eryilmaz et.~al.~\cite{Eryilmaz06} proposed another rate controller, so called \textit{Primal-Dual Congestion Controller}, which responds to congestion feedback gradually, rather than instantaneously. The gradual response is desired because the rate fluctuations are small. For instance, adaptive window flow control mechanisms such as transmission control protocol (TCP) respond to feedback gradually. 

The subgradient of $h(\mathbf{\lambda}, \mathbf{\mu})$ with respect to dual variable $\lambda_{i,j}$ is given by 
\begin{equation}\label{eq:subgradient-lambda}
\frac{\partial h}{\partial \lambda_{i,j}} = {\hat{c}}_{i,j} -\sum_{s\in\mathcal{S}}\sum_{k:(i,j)\in{p_{sk}}}{\Big(t_{sk}^{(i,j)}x_{sk}\Big)} -\sigma_{i,j}, \forall{(i,j)} 
\end{equation}

The subgradient of $h(\mathbf{\lambda}, \mathbf{\mu})$ with respect to dual variable $\mu_{sk}$, which is related to source node $s$ and its respective routing path $p_{sk}$, is expressed as 
\begin{equation}\label{eq:subgradient-mu}
\frac{\partial h}{\partial \mu_{sk}} = \mathcal{D}_{s} - \sum_{(i,j)\in\mathcal{L}}{\phi_{i,j}(R_{s})_{k(i,j)}}, \forall{p_{sk}\in \mathcal{P}_{s}}, \forall{s}
\end{equation}

In order to solve the dual problem of Eq.~\eqref{eq:dualopt}, we use a subgradient descent iteration method~\cite{Ber99} to update at each iteration $n$ the dual variables (Lagrangian multipliers) as follows
\vspace{-5pt}
\begin{multline}
%\begin{small}
\label{eq:lambda-update}
\lambda_{i,j}^{(n+1)} = \Bigg\{\lambda_{i,j}^{(n)}-\gamma\Bigg({\hat{c}}_{i,j}^{(n)} -  \\
 -\sum_{s\in\mathcal{S}}\sum_{k:(i,j)\in{p_{sk}}}{\Big(t_{sk}^{(i,j)}x_{sk}^{(n)}\Big)} -\sigma_{i,j}^{(n)}\Bigg)\Bigg\}^{+}
% \end{small}
\end{multline}
\vspace{-5pt}
%\end{small}
\begin{multline}
\label{eq:mu-update}
\mu_{sk}^{(n+1)} = \Bigg\{\mu_{sk}^{(n)} -\gamma\Bigg(\mathcal{D}_{s}  -\sum_{(i,j)\in\mathcal{L}}{\phi({\sigma_{i,j}^{(n)}})(R_{s})_{k(i,j)}}\Bigg)\Bigg\}^{+},
\end{multline}
where $\gamma$ is a positive step-size that ensures convergence of the iterative solution (e.g. $\gamma = 0.01$) and $(v)^{+} = \max(0,v)$ is the projection to the non-negative value. 
%According to Theorem $2.3$ in \cite{Shor}, 
%we know that there is no \textit{duality gap} for our problem and hence the dual optimization problem solution leads to the primal optimization problem optimal solution.

Based on the primal and dual variable updates of Eq.~\eqref{eq:rate_control_optimal},~\eqref{eq:delay_control_opt},~\eqref{eq:scheduling_opt},~\eqref{eq:lambda-update} and~\eqref{eq:mu-update}, we propose a distributed optimization algorithm, described below in Alg.~\ref{alg:distributedalg}.

\begin{algorithm}
\caption{Distributed Cross-Layer Optimization}\label{alg:distributedalg}
\begin{algorithmic}[1]
\STATE \textbf{INITIALIZE} primal and dual variables
\WHILE {$\mathbf{1}^{T}|\mathbf{x_{s}^{(n)}}-\mathbf{x_{s}^{(n-1)}}|\leq {\epsilon}$}
\STATE \textit{Dual Variables Update}\\
\STATE Each link $(i,j)$ updates its dual variable $\lambda_{i,j}$ (Eq.~\eqref{eq:lambda-update}).\\
%$\lambda_{i,j}^{(n+1)} = \Bigg\{\lambda_{i,j}^{(n)}-\gamma\Bigg({\hat{c}}_{i,j}^{(n)} -$  \\
% $-\underset{s\in\mathcal{S}}\sum\underset{k:(i,j)\in{p_{sk}}}\sum{\Big(x_{sk}^{(n)}t_{sk}^{(i,j)}\Big)} -\sigma_{i,j}^{(n)}\Bigg)\Bigg\}^{+}$
\STATE Each source $s$ updates the dual variables $\mu_{sk}$ (Eq.~\eqref{eq:mu-update}).\\
%$\mu_{sk}^{(n+1)} = \Bigg\{\mu_{sk}^{(n)} -\gamma\Bigg(\mathcal{D}_{s}  -\underset{(i,j)\in\mathcal{L}}\sum{\phi_{i,j}^{(n)}(R_{s})_{k(i,j)}}\Bigg)\Bigg\}^{+}$
\STATE \textit{Sources exchange dual variables}
\STATE Each source $s$ evaluates $\lambda^{s ,(n)}$.
\STATE Each source $s$ computes its traffic rate vector $\mathbf{x_{s}^{(n)}}$ by solving Eq.~\eqref{eq:rate_control_optimal}.
%$\mathbf{x_{s}^{(n+1)}} = \mathcal{U}_{s}^{'-1}\Big((\lambda^{s,(n)})^{T}\mathcal{T}_{s}^{T}\Big)$
\STATE Each link $(i,j)$ evaluates $\mu^{(i,j) ,(n)}$.
\STATE Each link $(i,j)$ computes its  $\mathbf{\sigma_{i,j}^{(n)}}$ by solving Eq.~\eqref{eq:delay_control_opt}.
\\ %$\mathbf{\sigma_{i,j}^{(n+1)}} = \Big[\mathcal{\phi}^{'-1}\Big(-\frac{\lambda_{i,j}^{(n)}}{\mu^{(i,j),(n)}}\Big)\Big]^{+}$
\STATE Each node performs scheduling via Eq.~\eqref{eq:scheduling_opt} as in~\cite{Delay-Constraints-NUM}.%Alg.~\ref{alg:scheduling}.
\ENDWHILE
\end{algorithmic}
\end{algorithm}

\section{Simulation Results\label{sec:eval}}
\begin{figure}[h]
    \centering
            \includegraphics[width=0.4\textwidth]{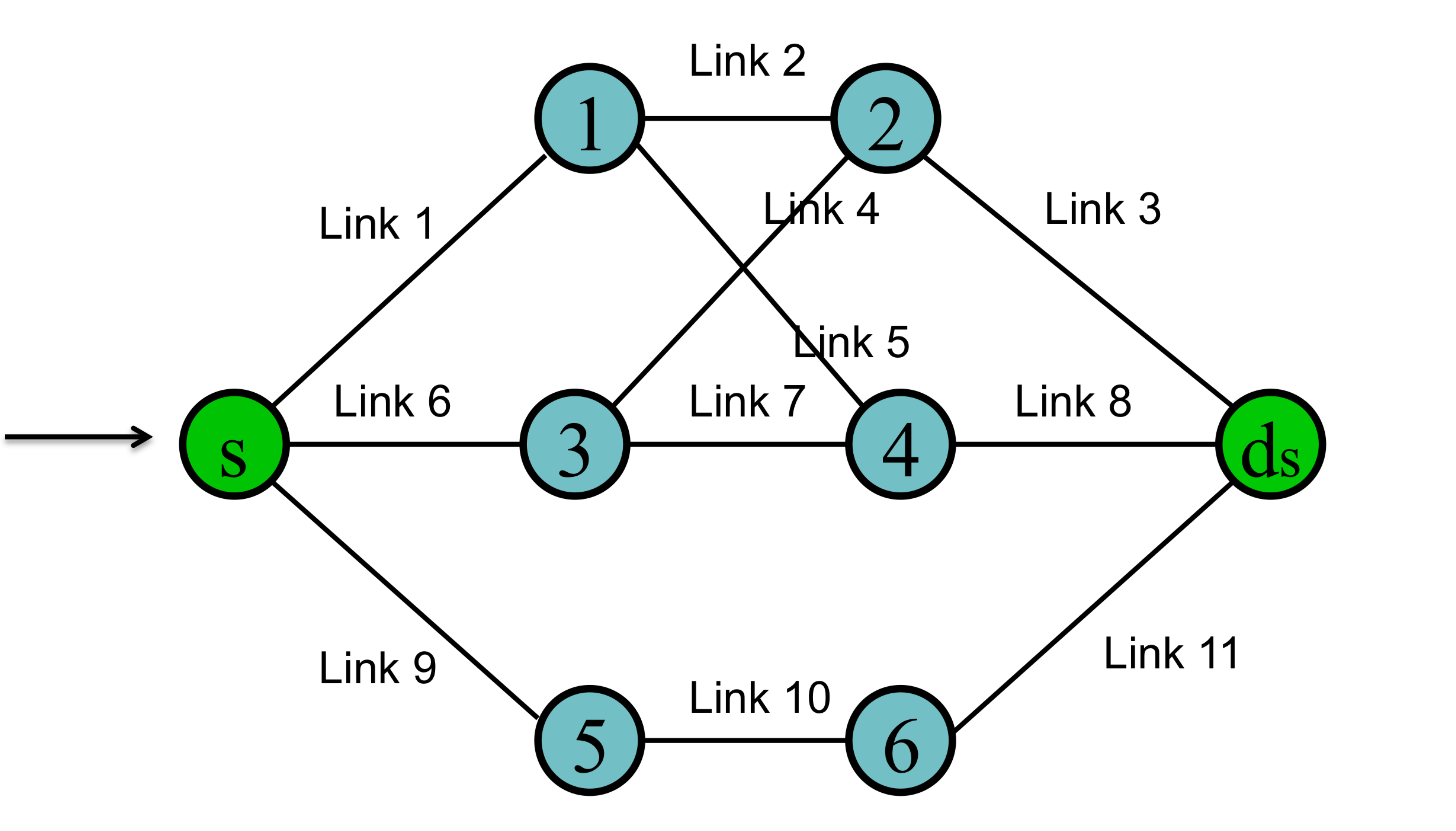}
            \caption{Wireless Network Scenario}
            \label{fig:SampleNetwork}
%\vspace{-5pt}
\end{figure}

In this section, we present simulation results for our trust-aware network utility maximization problem. 
 Fig.~\ref{fig:SampleNetwork} represents the sample wireless network scenario. The wireless network contains $\mathcal{N} = 8$ nodes and $\mathcal{L} = 11$ links, with maximum allowable capacity $c_{i,j}$ chosen in $[9,11]$ \textit{kbps}. There is one traffic flow from $s$ to $d_{s}$, which allocates traffic to different routing paths. Our simulation time is $T = 160$ time slots. The end-to-end delay constraint for the traffic flow is $\mathcal{D}_{s} = 2$ \textit{msec}. 
%We used CVX software~\cite{cvx} for our simulations.
% \vspace{-1pt}

 There are five different paths, where the source node $s$ can allocate data traffic to send to destination node $d_{s}$. The paths $p_{sk}$ are
 %$p_{s1}  = \{(s,1), (1,2), (2,d_{s})\}, p_{s2} =  \{(s,1), (1,4), (4,d_{s})\}, p_{s3} =  \{(s,3), (3,4), (4,d_{s})\}, p_{s4} =  \{(s,3), (2,4), (4,d_{s})\}$ and $p_{s5} =  \{(s,5), (5,6), (6,d_{s})\}$.
 %\vspace{-3pt}
 \begin{align*}
	p_{s1} & = \{(s,1), (1,2), (2,d_{s})\} \\
	p_{s2} &=  \{(s,1), (1,4), (4,d_{s})\} \\
	p_{s3} &=  \{(s,3), (3,2), (2,d_{s})\} \\
	p_{s4} &=  \{(s,3), (3,4), (4,d_{s})\} \\
	p_{s5} &=  \{(s,5), (5,6), (6,d_{s})\} 
\end{align*}
%\vspace{-3pt}

We define four trust \textit{update periods} (each period is defined every $T_{update}$ time slots), in order to show the behavior of our approach for different trust values. For the simulations, we define $T_{update} = T/4 = 40$ time slots. Node trust estimates $\nu_{i}$ change dynamically at every update period, based on the trust evaluation mechanism. The different node trust values that we obtain from the trust evaluation mechanism for each of the four update periods are shown at the matrix below

%\vspace{-10pt}
\begin{equation}
\label{eq:sample-trust-values}
\mathbf{\nu} = 
 \bordermatrix{ 
  ~ & s & \text{1} & \text{2} & \text{3} & \text{4} & \text{5} & \text{6} & d_{s} \cr  
  &1 & 1 & 1 & 0.7 & 1 & 0.7 & 0.5 &1 \cr
  &1 & 0.9 & 0.9 & 0.5 & 0.9 &  0.2 & 0.2 &1 \cr
  &1 & 0.9 & 0.9 & 0.3 & 0.7 & 0.1 & 0.1 &1 \cr
  &1 & 0.9 & 0.9 & 0.2 & 0.5 & 0.1 & 0.1 &1 \cr
}
 \end{equation}
 
 Trust values are adjusted using the EWMA algorithm expressed in Eq.~\eqref{eq:EWMA-trust}, in order to prevent significant variations in the trust estimates over subsequent trust update periods. For our simulation, the EWMA algorithm uses $\alpha = 0.8$ to give more significance to the latest update.
 
Given the trust values estimates in Matrix~\eqref{eq:sample-trust-values}, we can notice that path $p_{s5}$ contains untrusted (malicious) nodes and should ideally be excluded from the traffic rate assignment. In addition, node $3$ is detected to be malicious and hence our mechanism should ideally assign significantly less traffic to the paths $p_{s3}$ and $p_{s4}$ that contain this node. Finally, node $4$ obtains a low trust value estimate at the last update period, which should lead to decrease in the traffic rate assignment even for path $p_{s2}$ that contains this node.
 
\begin{figure}
	\centering
	\begin{subfigure}[b]{0.5\textwidth}
    \includegraphics[width=\textwidth]{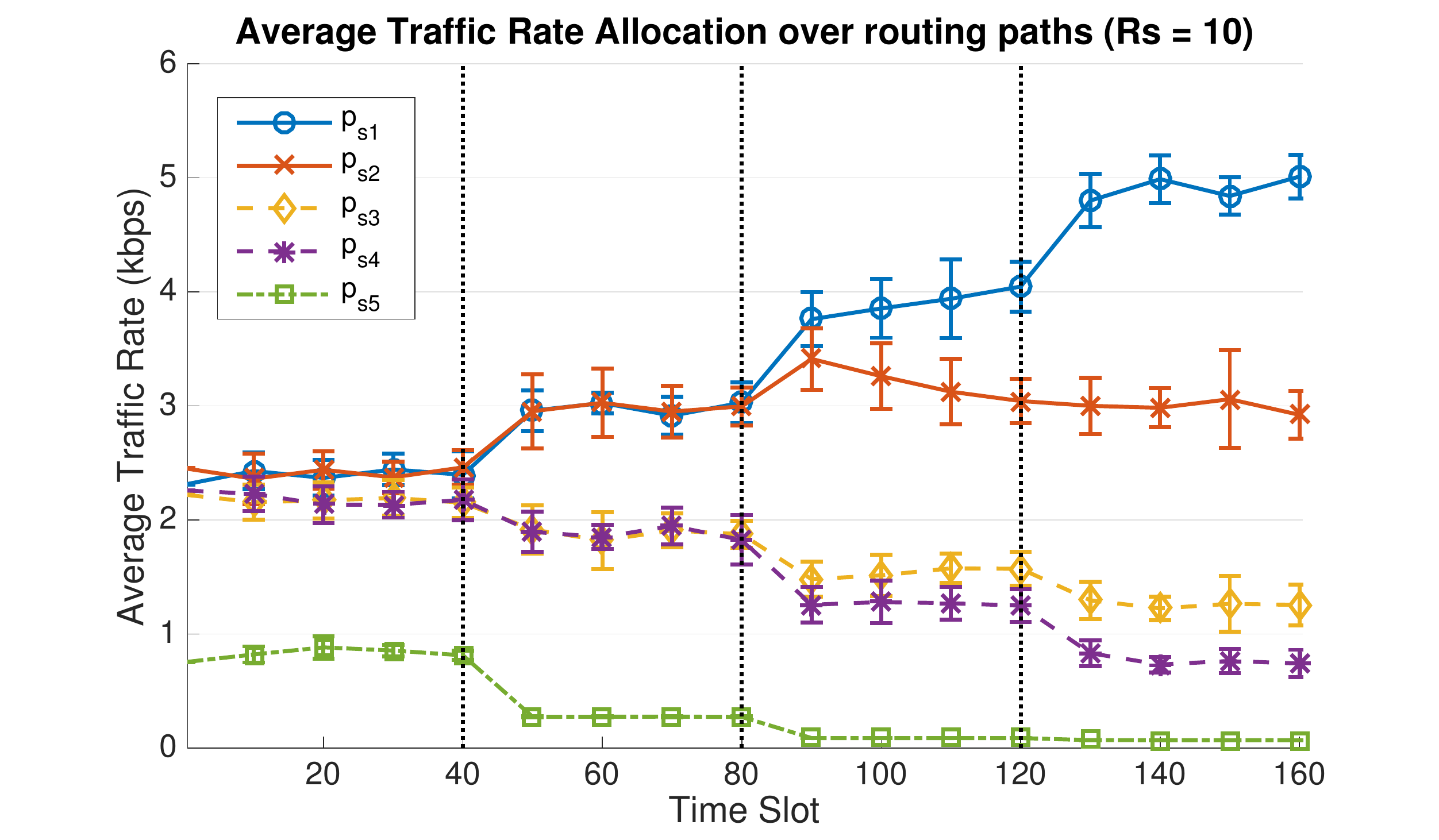}
     \caption{Average Traffic Rate with $\mathcal{R}_{s}= 10$}
      \label{fig:TrafficFlow10Slots160delay2}
\end{subfigure}
\vspace{20pt}
\begin{subfigure}[b]{0.5\textwidth}
			\includegraphics[width=\textwidth]{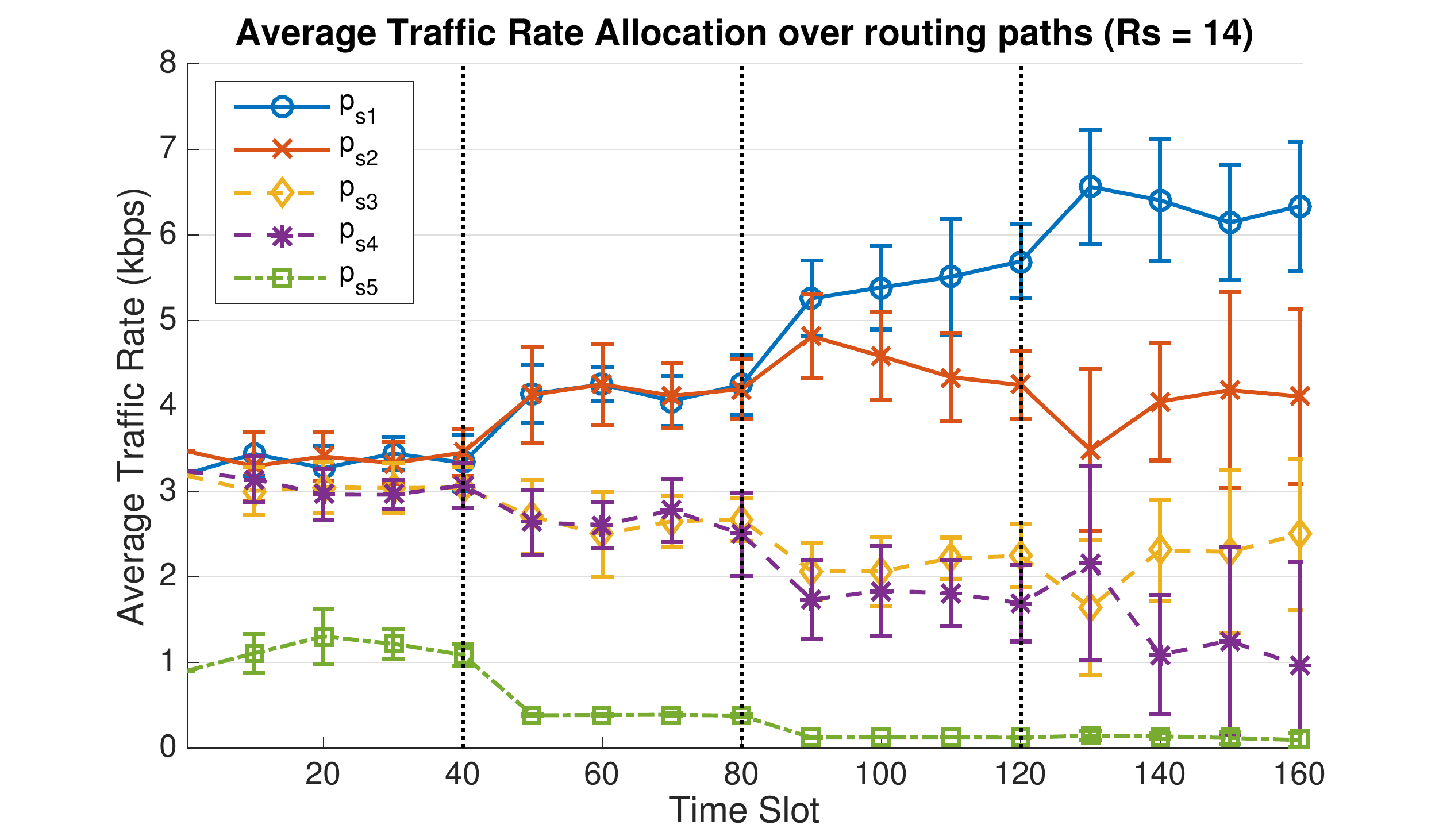}
            \caption{Average Traffic Rate with $\mathcal{R}_{s}= 14$}
            \label{fig:TrafficFlow14Slots160delay2}
\end{subfigure}
%\vspace{-12pt}
\caption{Average Traffic Rate over paths for different maximum rates $\mathcal{R}_{s}$}
\label{fig:AverageTrafficFlow}
%\vspace{-18pt}
\end{figure}

 Figures~\ref{fig:TrafficFlow10Slots160delay2} and~\ref{fig:TrafficFlow14Slots160delay2} present the numerical results of the average traffic rate allocation for two different cases of maximum allowable traffic rate $\mathcal{R}_{s}$ (with the corresponding error bars). In the case of $\mathcal{R}_{s} = 10$ \textit{kbps}, the maximum traffic rate is close to the maximum allowable capacity of the wireless links, while in the case of $\mathcal{R}_{s} = 14$ \textit{kbps}, the maximum traffic rate is greater than the maximum capacity of the links. We observe that in both cases the traffic rate assigned to each routing path changes at every update period based on the trust estimates. Our algorithm assigns to the path $p_{s1}$ the maximum traffic rate, since it contains trusted nodes and to the path $p_{s5}$ the lowest traffic rate, because it consists of untrusted nodes. For the rest of the paths, the traffic rate is being adjusted according to trust estimates of every update period. We also observe that in the case of $\mathcal{R}_{s} = 14$, more traffic rate is allocated to untrusted paths to cover the demand. 

\begin{figure}
	\centering
	\begin{subfigure}[b]{0.5\textwidth}
   \includegraphics[width=\textwidth]{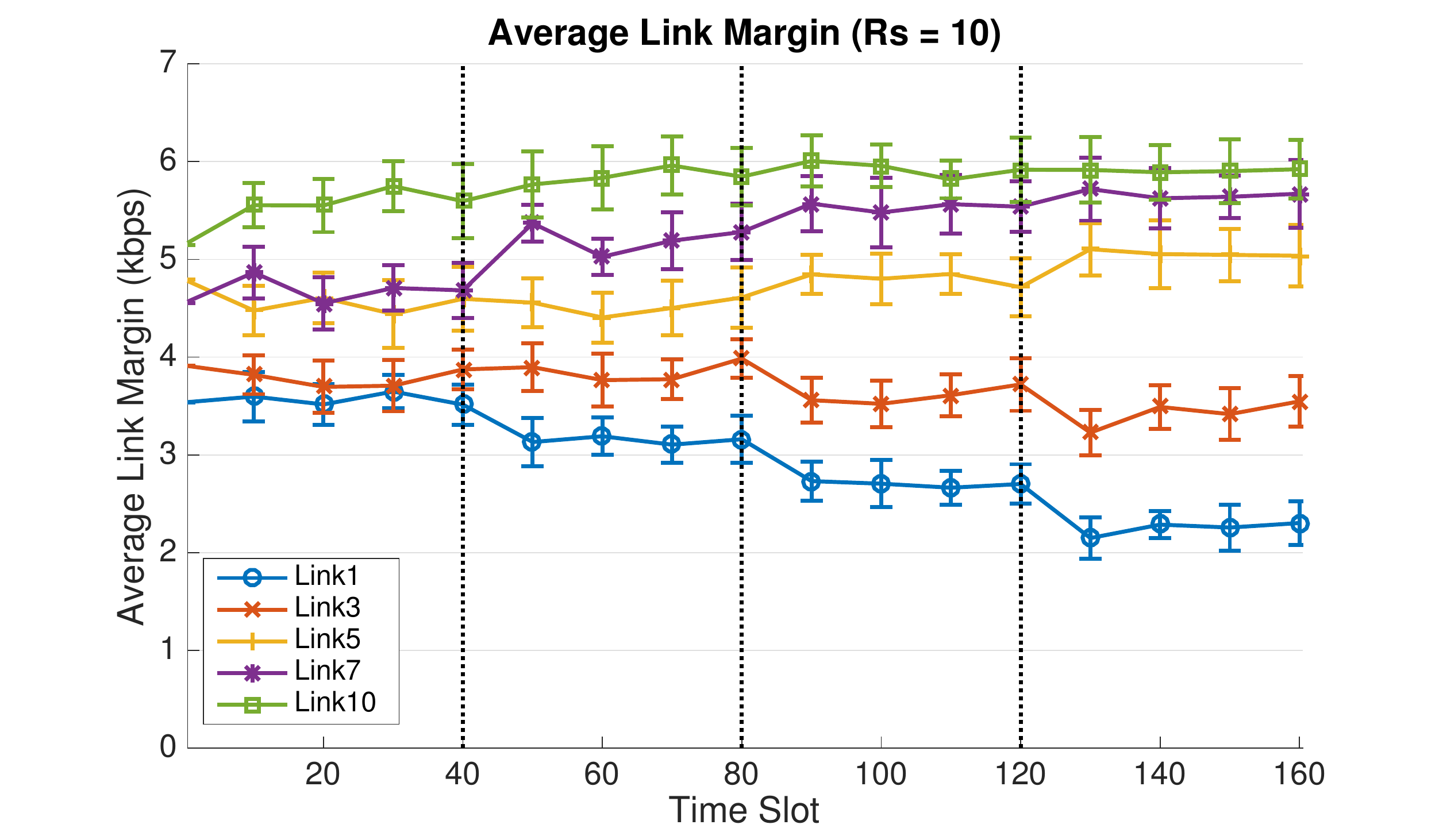}
            \caption{Average Link Margin with $\mathcal{R}_{s}= 10$}
            \label{fig:LinkMarginFlow10Slots160delay2}
\end{subfigure}
\vspace{20pt}
\begin{subfigure}[b]{0.5\textwidth}
		\includegraphics[width=\textwidth]{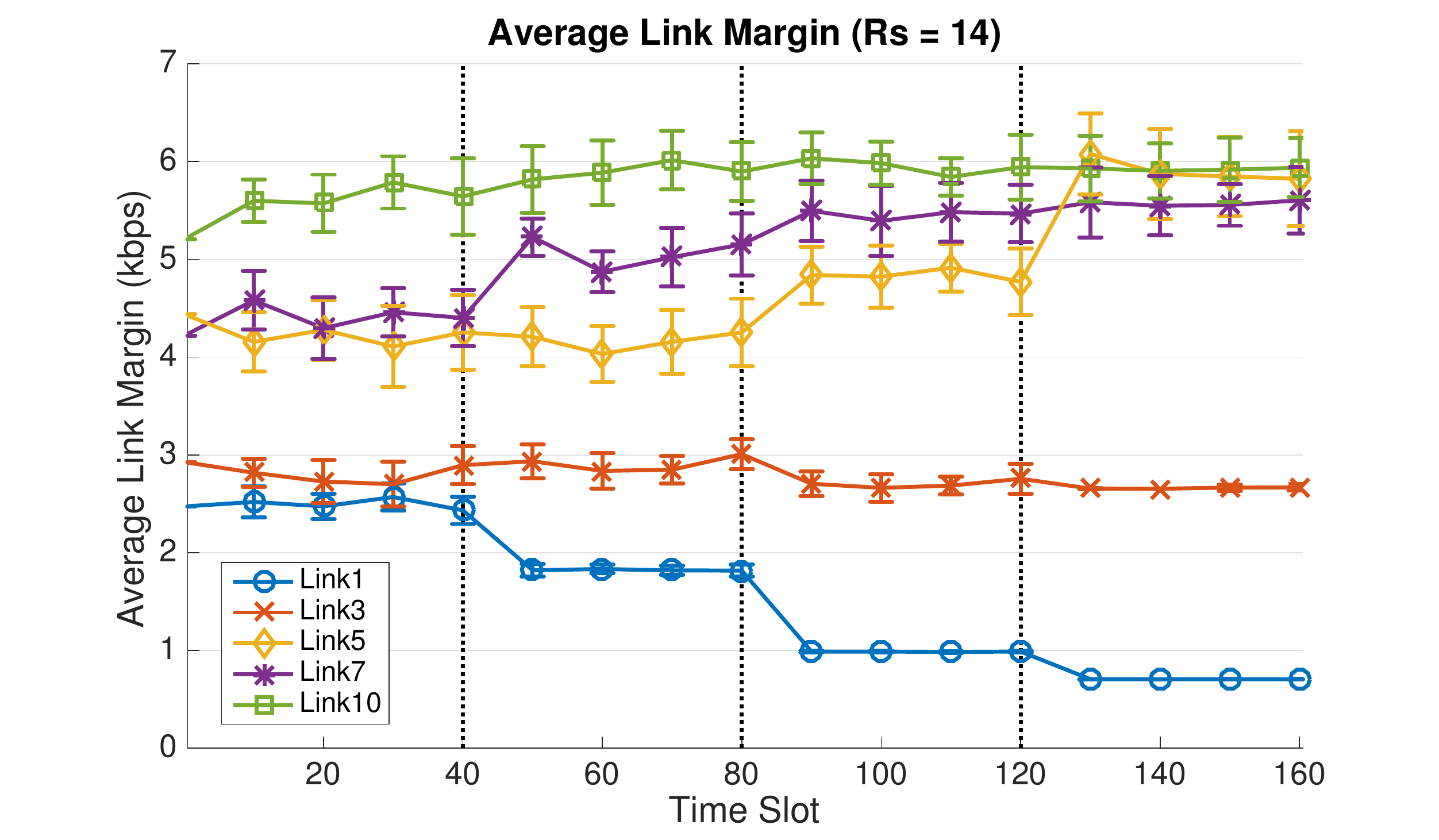}
		    \caption{Average Link Margin with $\mathcal{R}_{s}= 14$}
            \label{fig:LinkMarginFlow14Slots160delay2}
\end{subfigure}
%\vspace{-12pt}
\caption{Average Link Margin in wireless links over time}
\label{fig:AverageLinkMargin}
%\vspace{-18pt}
\end{figure}

Link capacity margins $\sigma_{i,j}$ for some links of our wireless network, in the case of $\mathcal{R}_{s} = 10$ and $\mathcal{R}_{s} = 14$, are presented in Figure~\ref{fig:LinkMarginFlow10Slots160delay2} and Figure~\ref{fig:LinkMarginFlow14Slots160delay2} respectively. Link capacity margin is related with the average delay, since higher capacity margin indicates lower link delay and thus lower end-to-end delay. In our scenario, link $1$ has the lowest capacity margin, because our scheme allocates significantly high traffic rate to this link. In addition, we notice that links belonging to untrusted paths have high capacity link margin (e.g. link $7$ and $11$), since they do not relay high data traffic. We also observe that in the case of higher maximum data rate, the link margin takes lower values at the more congested links, because of the higher traffic rate that should be allocated. Links that belong to untrusted paths, such as link $5$, are being allocated more traffic rate in the case of higher data rate, in order to satisfy the underlying delay requirements. In general, the capacity margin is being adjusted in order to attain the delay constraints of the traffic flow, which in our scenario are being satisfied, even if our scheme has to reduce significantly the capacity margin of some wireless links.

\section{Conclusion and Future Work\label{sec:conc}}
In this paper, we investigated an important application of performance and security tradeoff
by introducing security considerations in the cross layer
design of network protocols via network utility maximization (NUM).
The specific concept of security we used is \textit{“trust”}. Users get
higher utility by transmitting data through nodes of higher
trust values. Thus, trust values should be taken into account as parameters in
the optimization problem, so that the resulting trust-aware protocols are resilient
to network failures and to possible attacks. We also incorporated delay constraints in the utility optimization problem to capture QoS requirements. Finally, we proposed a distributed algorithm that achieves network utility
maximization. As part of future work, we plan to investigate how dynamic changes in trust values affect the utility optimization problem, and to evaluate our approach in large scale scenarios.

\bibliographystyle{IEEEtran.bst}
\bibliography{bib}

\begin{thebibliography}{10}
\providecommand{\url}[1]{#1}
\csname url@samestyle\endcsname
\providecommand{\newblock}{\relax}
\providecommand{\bibinfo}[2]{#2}
\providecommand{\BIBentrySTDinterwordspacing}{\spaceskip=0pt\relax}
\providecommand{\BIBentryALTinterwordstretchfactor}{4}
\providecommand{\BIBentryALTinterwordspacing}{\spaceskip=\fontdimen2\font plus
\BIBentryALTinterwordstretchfactor\fontdimen3\font minus
  \fontdimen4\font\relax}
\providecommand{\BIBforeignlanguage}[2]{{%
\expandafter\ifx\csname l@#1\endcsname\relax
\typeout{** WARNING: IEEEtran.bst: No hyphenation pattern has been}%
\typeout{** loaded for the language `#1'. Using the pattern for}%
\typeout{** the default language instead.}%
\else
\language=\csname l@#1\endcsname
\fi
#2}}
\providecommand{\BIBdecl}{\relax}
\BIBdecl

\bibitem{Chiang07}
M.~Chiang, S.~H. Low, A.~R. Calderbank, and J.~C. Doyle, ``Layering as
  optimization decomposition: A mathematical theory of network architectures,''
  \emph{Proceedings of the {IEEE}}, vol.~95, no.~1, pp. 255--312, January 2007.

\bibitem{ChiangINFOCOM06}
L.~Chen, S.~Low, M.~Chiang, and J.~Doyle, ``Cross-layer congestion control,
  routing and scheduling design in ad hoc wireless networks,'' in
  \emph{Proceedings of the 25th IEEE International Conference on Computer
  Communications (INFOCOM)}, April 2006, pp. 1--13.

\bibitem{StaiTON}
E.~Stai, S.~Papavassiliou, and J.~Baras, ``Performance-aware cross-layer design
  in wireless multihop networks via a weighted backpressure approach,''
  \emph{IEEE/ACM Transactions on Networking}, vol.~PP, no.~99, pp. 1--1, 2014.

\bibitem{Tassiulas92}
L.~Tassiulas and A.~Ephremides, ``Stability properties of constrained queueing
  systems and scheduling policies for maximum throughput in multihop radio
  networks,'' \emph{{IEEE} Transaction on Automatic Control}, vol.~37, no.~12,
  pp. 1936--1949, December 1992.

\bibitem{Lin06}
X.~Lin, N.~B. Shroff, and R.~Srikant, ``A tutorial on cross layer optimization
  in wireless networks,'' \emph{{IEEE} Journal on Selected Areas in
  Communications}, vol.~24, no.~8, pp. 1452--1463, August 2006.

\bibitem{Theodorakopoulos06}
G.~Theodorakopoulos and J.~Baras, ``On trust models and trust evaluation
  metrics for ad hoc networks,'' \emph{IEEE Journal on Selected Areas in
  Communications}, vol.~24, no.~2, pp. 318--328, Feb 2006.

\bibitem{Marti00}
S.~Marti, T.~J. Giuli, K.~Lai, and M.~Baker, ``Mitigating routing misbehavior
  in mobile ad hoc networks,'' in \emph{Proceedings of the 6th Annual
  International Conference on Mobile Computing and Networking}.\hskip 1em plus
  0.5em minus 0.4em\relax Boston, Massachusetts, United States: ACM Press,
  2000, pp. 255--265.

\bibitem{Delay-Constraints-NUM}
F.~Qiu, J.~Bai, and Y.~Xue, ``Towards optimal rate allocation in multi-hop
  wireless networks with delay constraints: A double-price approach,'' in
  \emph{Proceedings of the IEEE International Conference on Communications
  (ICC)}, June 2012, pp. 5280--5285.

\bibitem{Palomar06}
D.~P. Palomar and M.~Chiang, ``A tutorial on decomposition methods for network
  utility maximization,'' \emph{{IEEE} Journal on Selected Areas in
  Communications}, vol.~24, no.~8, pp. 1439--1451, August 2006.

\bibitem{Trichakis08dynamicnetwork}
N.~Trichakis, A.~Zymnis, and S.~Boyd, ``Dynamic network utility maximization
  with delivery contracts,'' in \emph{in Proceedings of the IFAC World
  Congress}, 2008, pp. 2907--2912.

\bibitem{DBLP:journals/corr/HajiesmailiTK15}
M.~H. Hajiesmaili, M.~S. Talebi, and A.~Khonsari, ``Utility-optimal dynamic
  rate allocation under average end-to-end delay requirements,'' \emph{CoRR},
  vol. abs/1509.03374, 2015.

\bibitem{TaoISCCSP}
J.~Baras, T.~Jiang, and P.~Purkayastha, ``Constrained coalitional games and
  networks of autonomous agents,'' in \emph{Proceedings of the 3rd
  International Symposium on Communications, Control and Signal Processing
  (ISCCSP)}, March 2008, pp. 972--979.

\bibitem{Jamming-Aware-NUM}
P.~Tague, S.~Nabar, J.~Ritcey, and R.~Poovendran, ``Jamming-aware traffic
  allocation for multiple-path routing using portfolio selection,''
  \emph{IEEE/ACM Transactions on Networking}, vol.~19, no.~1, pp. 184--194, Feb
  2011.

\bibitem{Sherry:2015:BDP:2785956.2787502}
J.~Sherry, C.~Lan, R.~A. Popa, and S.~Ratnasamy, ``Blindbox: Deep packet
  inspection over encrypted traffic,'' in \emph{Proceedings of the 2015 ACM
  Conference on Special Interest Group on Data Communication}, ser. SIGCOMM,
  New York, NY, USA, pp. 213--226.

\bibitem{Blaze99}
M.~Blaze, J.~Feigenbaum, J.~Ioannidis, and A.~D. Keromytis, ``The role of trust
  management in distributed systems security,'' \emph{Secure Internet
  Programming: Security Issues for Mobile and Distributed Objects}, pp.
  185--210, 1999.

\bibitem{Eschenauer02ontrust}
L.~Eschenauer, V.~D. Gligor, and J.~Baras, ``On trust establishment in mobile
  ad-hoc networks,'' in \emph{Proceedings of the Security Protocols
  Workshop}.\hskip 1em plus 0.5em minus 0.4em\relax Springer-Verlag, 2002, pp.
  47--66.

\bibitem{JiangGlobecom08}
T.~Jiang and J.~Baras, ``Trust credential distribution in autonomic networks,''
  in \emph{Proceedings of IEEE Global Telecommunications Conference
  (GLOBECOM)}, Nov 2008, pp. 1--5.

\bibitem{Maurer:1996:MPI:646646.699185}
U.~M. Maurer, ``Modelling a public-key infrastructure,'' in \emph{Proceedings
  of the 4th European Symposium on Research in Computer Security: Computer
  Security (ESORICS)}, London, UK, 1996, pp. 325--350.

\bibitem{doi:10.1080/00401706.2000.10485986}
S.~W. Roberts, ``Control chart tests based on geometric moving averages,''
  \emph{Technometrics}, vol.~42, no.~1, pp. 97--101, 2000.

\bibitem{Rate-ReliabilityTradeoff}
J.-W. Lee, M.~Chiang, and A.~Calderbank, ``Price-based distributed algorithms
  for rate-reliability tradeoff in network utility maximization,'' \emph{IEEE
  Journal on Selected Areas in Communications}, vol.~24, no.~5, pp. 962--976,
  May 2006.

\bibitem{Jain:2003:IIM:938985.938993}
K.~Jain, J.~Padhye, V.~N. Padmanabhan, and L.~Qiu, ``Impact of interference on
  multi-hop wireless network performance,'' in \emph{Proceedings of the 9th
  Annual International Conference on Mobile Computing and Networking
  (MobiCom)}, New York, NY, USA, 2003, pp. 66--80.

\bibitem{Boyd04}
S.~Boyd and L.~Vandenberghe, \emph{Convex Optimization}.\hskip 1em plus 0.5em
  minus 0.4em\relax New York, NY, USA: Cambridge University Press, 2004.

\bibitem{Ber99}
D.~P. Bertsekas, \emph{Nonlinear Programming}.\hskip 1em plus 0.5em minus
  0.4em\relax Belmont, MA: Athena Scientific, 1999.

\end{thebibliography}

%\begin{thebibliography}{1}
%\end{thebibliography}

\end{document}